\documentclass[a4paper]{jpconf}
\usepackage{graphicx}
\usepackage{amsbsy}
\begin{document}
\title{Cosmological perturbations from multi-field inflation}
\author{David Langlois}
\address{{\small {\it APC (Astroparticules et Cosmologie),}}\\
{\small {\it UMR 7164 (CNRS, Universit\'e Paris 7, CEA, Observatoire de
Paris)}}\\
{\small {\it  10, rue Alice Domon et L\'eonie Duquet,
 75205 Paris Cedex 13, France}}\\
 {\small {\it and}}\\
 {\small {\it Institut d'Astrophysique de Paris (IAP),}}\\
{\small {\it 98bis Boulevard Arago, 75014 Paris, France}}}

\def\half{\frac{1}{2}}
\def\beq{\begin{equation}}
\def\eeq{\end{equation}}
\newcommand{\bea}{\begin{eqnarray}}
\newcommand{\eea}{\end{eqnarray}}
\def\Tdot#1{{{#1}^{\hbox{.}}}}
\def\Tddot#1{{{#1}^{\hbox{..}}}}
\def\p{\phi}
\newcommand{\dn}[2]{{\mathrm{d}^{{#1}}{{#2}}}}
\def\PX{P_{,X}}
\def\s{\sigma}
\def\mP{m_P}
\def\R{{\cal R}}
\def\tX{{\tilde X}}
\def\tG{{\tilde G}}
\def\tP{{\tilde P}}

\begin{abstract}
We briefly review  the standard derivation of the spectra of cosmological perturbations in the simplest
models of inflation. We then consider models with several scalar fields, described by  Lagrangians with an arbitrary dependence on the  kinetic terms. We  illustrate our general formalism with the case of multi-field DBI inflation.
\end{abstract}

\section{Introduction}

Inflation has now become a standard paradigm to describe the physics of the very early universe, and as cosmological data keep improving, one can hope to get more and more clues about the very early universe. So far, the simplest models of inflation are compatible with the data but it is instructive to study more refined models for at least two reasons. First, because models inspired by high energy physics are usually more complicated than the simplest phenomenological inflationary models. Second, because these generalized models will  give us an idea of how much the future data will be able to pin down some specific region in parameter space. 

In the first part of this contribution, we present a brief summary of the standard approach to compute the cosmological perturbations in the simplest inflationary models. This summary is mainly based on the pedagogical introduction \cite{cargese} where the reader will find more details.

In the second part, we show how the standard results are modified when several scalar fields play a role during inflation. We present  recent results for very general multi-field inflationary models, allowing for non-standard kinetic terms. This generalization is  motivated by efforts  to connect  string theory and inflation and we focus our attention on multi-field DBI (Dirac-Born-Infeld) inflation.

\section{Standard single field inflation}
In this section, we consider the simplest inflationary models, which are based on  a single scalar field $\phi$ governed by an 
action of the form
\beq
S=\int d^4x\sqrt{-g}\left(-{1\over 2}\partial^\mu\phi\partial_\mu\phi
-V(\phi)\right),
\eeq
where $V(\phi)$ is the potential for the scalar field.
The corresponding energy-momentum tensor is given by  
\beq
T_{\mu\nu}=\partial_\mu\phi\partial_\nu\phi-g_{\mu\nu}
\left({1\over 2}\partial^\sigma\phi\, \partial_\sigma\phi
+V(\phi)\right).
\label{Tscalarfield}
\eeq
In a FLRW (Friedmann-Lema\^\i tre-Robertson-Walker) spacetime, with metric
\beq
ds^2=-dt^2+a^2(t) d{\vec x}^2,
\eeq
 the energy-momentum tensor reduces to the perfect fluid form
with  energy density and pressure given respectively by
\beq
\rho=-T_0^0={1\over 2}\dot\phi^2+V(\phi),\quad 
p={1\over 2}\dot\phi^2-V(\phi).
\eeq
The equation of motion for the scalar field is 
\beq
\label{KG}
\ddot\phi+3H\dot \phi+V'=0. 
\eeq
and the evolution of the scale factor is governed by  Friedmann's equations
\begin{eqnarray}
H^2={8\pi G\over 3}\left({1\over 2}\dot\phi^2+V(\phi)\right), \qquad  \dot H=-4\pi G\dot\phi^2.
\label{friedmann1}
\end{eqnarray}

If the potential satisfies the so-called slow-roll conditions,
\beq
\epsilon_V\equiv {\mP^2\over 2}\left({V'\over V}\right)^2 \ll 1, \quad \eta_V\equiv \mP^2 {V''\over V}\ll 1,
\eeq
where 
$m_P\equiv (8\pi G)^{-1/2}$ is the reduced Planck mass, the evolution can enter into a  {\it slow-roll inflationary regime}
where the kinetic energy of the scalar field  in (\ref{friedmann1}) 
and the acceleration $\ddot\phi$ in the  Klein-Gordon equation 
(\ref{KG}) are negligible.

In order to study the linear cosmological perturbations, one must perturb the matter, i.e. the scalar field, as well as the geometry, i.e. the metric. Restricting ourselves to scalar perturbations, the metric reads
\beq 
ds^2= -(1+2A)dt^2+ 2 a(t) \partial_iB\, dx^idt+
a^2(t)\left[(1-2\psi)\delta_{ij}
+2\partial_i\partial_jE\right]dx^idx^j,
\eeq
where $\psi$ is directly related to the intrinsic curvature of constant time hypersurfaces, according to the relation
\beq
{}^{(3)}R=\frac{4}{a^2}\nabla^2\psi.
\eeq
The metric perturbations are modified in a change of coordinates. It is thus useful (although not necessary) to define gauge-invariant coordinates, such as the curvature perturbation on uniform energy hypersurfaces,
\beq
-\zeta\equiv \psi+\frac{{ H}}{\dot\rho}\delta\rho
=\psi-\frac{\delta\rho}{3(\rho+p)},
\eeq
or 
the comoving curvature perturbation, 
\beq
{\cal R}=\psi-\frac{H}{\rho+p}\delta q \,, 
\eeq
where $\delta q$ is the scalar part of the momentum density ($\delta T_i^0\equiv \partial_i\delta q$). 
Using the linearized Einstein's equations, it can be shown that these two quantities are related via
\beq
\zeta=-{\cal R}-{2\rho\over 3(\rho+P)}\left({k\over aH}\right)^2\Psi
\eeq
where
\beq
\label{Psi}
\Psi = \psi + a^2 H (\dot{E}-B/a).
\eeq
The quantity $\zeta$ is particularly interesting because it can be shown to be constant on large scales when the matter perturbations are adiabatic, i.e. when they satisfy 
\beq
\delta P_{\rm nad}\equiv 
\delta p-{{\dot p}\over {\dot\rho}}\, \delta\rho=0.
\eeq
This property, which is well-known for linear perturbations, can be seen as the consequence of a more general 
result. Indeed, as shown  in \cite{Langlois:2005ii,Langlois:2005qp}, the conservation of the energy-momentum tensor 
for any perfect fluid, characterized by the energy density $\rho$, the pressure $p$ and the four-velocity $u^a$, leads 
to the {\it exact} relation
\beq
\label{dot_zeta}
\dot\zeta_a\equiv {\cal L}_u\zeta_a=
-{\Theta\over{3(\rho+p)}}\left( \nabla_a p -
\frac{\dot p}{\dot \rho} \nabla_a\rho\right), 
\eeq
where we have defined 
\beq
 \zeta_a\equiv 
\nabla_a\alpha-\frac{\dot\alpha}{\dot\rho}\nabla_a\rho, \quad \Theta=\nabla_a u^a, \quad \alpha=\frac{1}{3}\int d\tau \,
\Theta,
\eeq
and where a dot on scalar quantities denotes a derivative along $u^a$ (e.g. $\dot\rho\equiv u^a\nabla_a\rho$). This identity is valid for any spacetime geometry and does not rely on Einstein's equations. In the cosmological context, $\alpha$ can be interpreted as a non-linear generalization, according to an observer following the fluid, of the number of e-folds of the scale factor. Introducing an explicit coordinate system and linearizing (\ref{dot_zeta}) leads to the familiar result of the linear theory.

During inflation, it is easier to work with the perturbation $\R$, since in this case
\beq
{\cal R}=\psi+\frac{H}{\dot\phi} \delta\phi\, .
\eeq
Because of the constraints arising from Einstein's equations, the scalar metric perturbations and the scalar field perturbation are not independent. In fact, there is only one degree of freedom which can be expressed in terms of the combination
\beq
v=a\left(\delta\phi+\frac{\dot\phi}{H} \psi\right)\equiv a \, Q\, ,
\eeq
where $Q$ represents the scalar field perturbation in the spatially flat gauge (where $\psi=0$).
The quadratic action governing the dynamics of this degree of freedom can be obtained from the expansion up to second order of the full action. One finds
\beq
S_v={1\over 2}\int d\tau  \, \, d^3x\, \left[{v'}^2+\partial_i v \partial ^i v
+{z''\over z}v^2\right],
\eeq
where a prime denotes a derivative with respect to the conformal time $\tau=\int dt/a(t)$, and 
with
\beq
\label{z}
z=a\frac{\dot\phi}{H}.
\eeq
To quantize this system, one considers $v$ as a quantum field and one 
 decomposes it  as 
\beq
\label{Fourier_quantum}
\hat v (\tau, \vec x)={1\over (2\pi)^{3/2}}\int d^3k \left\{{\hat a}_{\vec k} v_{k}(\tau) e^{i \vec k.\vec x}
+ {\hat a}_{\vec k}^\dagger v_{k}^*(\tau) e^{-i \vec k.\vec x} \right\},
\eeq
where  the $\hat a^\dagger$ and  $\hat a$ are 
 creation and annihilation operators , which satisfy the 
usual commutation rules 
\beq
\label{a}
\left[ {\hat a}_{\vec k}, {\hat a^\dagger}_{\vec k'}\right]= \delta(\vec k-\vec k')\, ,
\quad
\left[ {\hat a}_{\vec k}, {\hat a}_{\vec k'}\right]= 
\left[ {\hat a^\dagger}_{\vec k}, {\hat a^\dagger}_{\vec k'}\right]= 0\, .
\eeq
The action  implies that  the conjugate momenta for $v$ is
$v'$. Therefore,  the canonical quantization  
for $\hat v$ and its conjugate momentum leads to  the condition
\beq
v_{k} {v'_{k}}^*-v_{k}^*v'_{k}=i\,.
\label{wronskien}
\eeq
The complex function  $v_{k}(\tau)$ satisfies the equation of motion 
\beq
v''+\left( k^2-\frac{z''}{z}\right) v=0.
\eeq
In the slow-roll limit,  $z''/z\simeq 2/\tau^2$, and one can use the solution for a de Sitter spacetime (where $H$ is constant). Note that this is only an approximation as the Hubble parameter is decreasing with time, but a very good one,  when the slow-roll parameters are small, during the short time when the scale of interest crosses out the Hubble 
radius ($k\sim a H$).
 Requiring that the solution on small scales behaves like the Minkowski vacuum selects  the particular solution 
\beq
v_{k}\simeq  \frac{1}{\sqrt{2k}}e^{-ik \tau }\left(1-{i\over k \tau}\right),
\eeq
where the normalization is imposed by the condition (\ref{wronskien}). 
This implies that the power spectrum of the scalar field fluctuations is given by
\beq
{\cal P}_{Q}=\frac{k^3}{2\pi^2}|v_k|^2\frac{1}{a^2}\simeq\frac{H^2}{4\pi^2},
\eeq
where the quantities on the right hand side are evaluated at 
 {\it Hubble crossing}. This can be translated into the power spectrum of the curvature perturbation 
 ${\cal R}$, by noting that ${\cal R}=aQ/z$. One thus gets
\beq
{\cal P}_{\cal R}=\frac{k^3}{2\pi^2}\frac{|v_{k}|^2}{z^2}\simeq\left(\frac{H^4}{4\pi^2 \dot\phi^2}\right)_{|k=aH}.
\eeq
In single-field inflation, since ${\cal R}$ is conserved on large scales (as ${\cal R}$ and $\zeta$ coincide on large scales), the above expression, evaluated at Hubble crossing, determines the amplitude 
of the curvature perturbation just before the modes reenter the Hubble radius and thus sets 
the initial conditions for cosmological perturbations. 

\section{Generalized multi-field inflation}
We now consider  multi-field models,  which can be described by an
action of the form \cite{lrst08b}
\beq
\label{P}
S =  \int d^4 x \sqrt{-g}\left[\frac{R}{16\pi G}   +   P(X^{IJ},\phi^K)\right] 
\eeq
where $P$ is an arbitrary function of $N$ scalar fields and of the  kinetic term
\beq
X^{IJ}=-\half \nabla_\mu \p^I  \nabla^\mu \p^J.
\eeq
The very general form (\ref{P})  can be seen as an extension of the Lagrangian  of k-inflation \cite{ArmendarizPicon:1999rj} to the case of  several scalar fields. 

A more restrictive class of models, considered in \cite{lr08}, consists of Lagrangians that depend on a global kinetic term 
$X=G_{IJ}X^{IJ}$ where $G_{IJ} \equiv G_{IJ}(\p^K)$ is an arbitrary metric on the $N$-dimensional field space.
By defining $P=X-V$, one recovers in particular multi-field models with an action of the form
\beq
S=\int d^4x\sqrt{-g}\left(-{1\over 2}\, 
G_{IJ}(\phi)
\, \partial^\mu\phi^I\partial_\mu\phi^J
-V(\phi)\right),
\eeq
where in the simplest cases, one can take a flat metric in field space ($G_{IJ}=\delta_{IJ}$), so that the kinetic terms are standard.

The relations obtained in the previous section for  the single field model can then be generalized.
The energy-momentum tensor, derived from (\ref{P}),   is of the form
\beq
T^{\mu \nu} = P g^{\mu \nu} + P_{<IJ>}  \partial^\mu \phi^I \partial^\nu \phi^J\,,
\label{Tmunu}
\eeq
where $P_{<IJ>}$ denotes the partial derivative of $P$ with respect to $X^{IJ}$ (symmetrized with respect to the indices $I$ and $J$).
The equations of motion for the scalar fields, which can be seen as generalized Klein-Gordon equations, are obtained from the variation of the action with respect to $\phi^I$. One finds
\beq
\nabla_{\mu} \left(  P_{<IJ>} \nabla^\mu \phi^J \right) + P_{,I} = 0\,.
\label{KG-general}
\eeq
where $P_{,I}$ denotes the partial derivative of $P$ with respect to $\phi^I$.

In a spatially flat FLRW (Friedmann-Lema\^itre-Robertson-Walker) spacetime, with metric
\beq
ds^2=-dt^2+a^2(t)d{\vec x}^2,
\eeq
the scalar fields are homogeneous, so that $X^{IJ}=\dot\phi^I\dot\phi^J/2$, and the energy-momentum tensor reduces to that of a perfect fluid
with energy density 
\beq
\rho=2  P_{<IJ>}  X^{IJ} - P\,,
\label{rho}
\eeq
and pressure $P$. 
 The evolution of the scale factor $a(t)$ is governed by the Friedmann equations, which can be written in the form
\beq
H^2 = \frac{1}{3} \left(2  P_{<IJ>}  X^{IJ} - P \right)\, , \qquad
\dot{H} = - X^{IJ} P_{<IJ>} \,.
\label{Friedmann2}
\eeq
The equations of motion for the scalar fields reduce to
\beq
\left( P_{<IJ>} + P_{<IL>,<JK>} \dot{\phi}^L \dot{\phi}^K  \right) \ddot{\phi}^J+ 
 \left( 3 H P_{<IJ>} + {P}_{<IJ>,K} \dot{\phi}^K\right)\dot{\phi}^J - P_{,I} = 0\, ,
\label{KG1}
\eeq
where $P_{<IL>,<JK>}$ denotes the (symmetrized) second derivative of $P$ with respect to $X^{IL}$ and $X^{JK}$.

The expansion up to second order in the linear perturbations of the action (\ref{P}) is useful to obtain the classical equations of motion for the perturbations. It is also the starting point to calculate the spectra of the primordial perturbations generated from the vacuum quantum fluctuations of the scalar fields during inflation, as we have seen in the previous section for a single scalar field.  
 Working for convenience with the scalar field perturbations 
$Q^I$ defined in the spatially flat gauge, the
 second order action can be  written in the rather simple form \cite{lrst08b}
 \begin{eqnarray}
S_{(2)} &=& \frac{1}{2} \int {\rm d}t \, {\rm d}^3x \, a^3 \left[ 
\left(P_{<IJ>} + 2 P_{<MJ>,<IK>}X^{MK}\right) \dot{Q}^{I}\dot{Q}^{J}  - P_{<IJ>} h^{ij} \partial_iQ^I \partial_jQ^J \right.
\\
&& \qquad \qquad \qquad \qquad \left. - {\cal M}_{KL}Q^K Q^L 
 + 2 \,\Omega_{KI}Q^K \dot{Q}^I  \right] \qquad
 \label{S2}
\end{eqnarray}
where the mass matrix is
\begin{eqnarray}
{\cal M}_{KL} &=& - P_{,KL} + 3 X^{MN} P_{<NK>} P_{<ML>} 
+ \frac{1}{H} P_{<NL>}\dot{\phi}^N \left[ 2 P_{<IJ>,K} X^{IJ} - P_{,K} \right]
\\
& - & \frac{1}{H^2} X^{MN} P_{<NK>}P_{<ML>} \left[ X^{IJ} P_{<IJ>} + 2 
P_{<IJ>,<AB>} X^{IJ}X^{AB} \right] 
\\
& - & { \frac{1}{a^3}\frac{{\rm d}}{{\rm dt}} } \left( \frac{a^3}{H} P_{<AK>} P_{<LJ>} X^{AJ} \right)
\label{masssq}
\end{eqnarray}
and the mixing matrix is
\beq
\Omega_{KI} = \dot{\phi}^J P_{<IJ>,K} -
 \frac{2}{H} P_{<LK>} 
2 P_{<MJ>,<NI>} X^{LN}X^{MJ}\,.
\label{damping}
\eeq

A particularly interesting model of the form (\ref{P}) is the multi-field extension of DBI (Dirac-Born-Infeld) inflation \cite{dbi}, which arises when one considers the motion of a D-brane in a warped throat while taking into account a possible angular motion. 
As shown in \cite{lrst08a}, the corresponding Lagrangian is of the form 
\beq
P(X^{IJ},\phi^K)= -\frac{1}{f(\phi^I)}\left(\sqrt{{\cal D}}-1\right) -V(\phi^I),
\label{DD1}
\eeq
with
\beq
\label{def_explicit}
{\cal D}=1-2f G_{IJ}X^{IJ}+4f^2 X^{[I}_IX_J^{J]} -8f^3 X^{[I}_IX_J^{J} X_K^{K]}+16f^4 X^{[I}_IX_J^{J} X_K^{K}X_L^{L]}
\equiv 1-2f \tX\, \, 
\eeq
where the field indices are lowered by the field metric $G_{IJ}$, which naturally comes from the higher-dimensional 
spacetime where the scalar fields $\phi^I$ correspond to the coordinates of the D-brane. 

It is convenient to rewrite the Lagrangian (\ref{DD1}) as a function of $\tX$, introduced just above, 
\beq
 P(X^{IJ},\phi^K) =  \tilde{P}(\tilde{X},\phi^K) = -\frac{1}{f(\phi^I)} \left(\sqrt{1-2f(\phi^I)\tilde{X}}-1 \right) - V(\phi^I).
\label{Ptilde}
\eeq
Note that $\tX$ and $X$ coincide in the homogeneous background. 
The situation is then very similar to Lagrangians of the form $P=P(X,\phi^K)$ where $X=G_{IJ}X^{IJ}$, studied in 
\cite{lr08}, and one can rewrite the second order action in terms of the covariant derivative $ \mathcal{D}_I$ defined with respect to the field space metric $G_{IJ}$. This gives \cite{lrst08a}
\begin{eqnarray}
S_{(2)}&=&\half \int {\rm d}t \,\dn{3}{x}\,    a^3\left[ \tP_{,X}\left( \tG_{IJ}
 \mathcal{D}_t Q^I \mathcal{D}_t Q^J   -\frac{c_s^2}{a^2}\tG_{IJ} \partial_i Q^I \partial^i Q^J \right)
 \right.
 \cr
  && 
  \left.
  - {\tilde {\cal M}}_{IJ}Q^I Q^J + 2 \tP_{,XJ} \dot \p_I Q^J \mathcal{D}_t Q^I  \right]\,,
\label{2d-order-action}
\end{eqnarray}
where we can  substitute $\tP_{,X}=1/c_s$ and $ \tP_{,X J}  ={f_J X}/{c_s^{3}}$. In (\ref{2d-order-action}),
we have  introduced the time covariant derivative $\mathcal{D}_t Q^{I} \equiv \dot{Q}^I + \Gamma^{I}_{JK} \dot{\phi}^J Q^{K}$  where $\Gamma^{I}_{JK} $ is the Christoffel symbol constructed from $G_{IJ}$  (and $\mathcal{R}_{IKLJ}$ will denote the corresponding Riemann tensor).
We have also introduced the {\it effective speed of sound}
\beq
c_s=\sqrt{1-2fX},
\eeq
which corresponds to the propagation speed of linear perturbations, as well as the  
 {\it deformed} field metric 
\beq
\label{Gtilde}
\tilde{G}_{I}^{\, J}=\perp_{I}^{\, J}+\frac{1}{c_s^2} e_I e^J
, \qquad \perp_{I}^{\, J}=\delta_{I}^{\, J}-e_Ie^J \, ,
\eeq
where
\beq
\label{e1}
e^I=\frac{\dot{\p^I}}{\sqrt{2X}},
\eeq
is the unit vector along the inflationary trajectory in field space. 
Finally the effective squared mass matrix which appears above, and which differs from ${\cal M}_{IJ}$ in Eq.~(\ref{masssq}), is
\begin{eqnarray}
{\tilde{\cal M}}_{IJ} &=& -\mathcal{D}_I \mathcal{D}_J \tP - \tP_{,\tX} \mathcal{R}_{IKLJ}\dot \p^K \dot
\p^L+\frac{\tX \tP_{,\tX}}{H} (\tP_{,\tX J}\dot \p_I+\tP_{,\tX I}\dot \p_J)+ \frac{\tX \tP_{,\tX}^3}{2 H^2}\left(1-\frac{1}{c_s^2}\right)\dot \p_I \dot \p_J
 \nonumber\\
 &&~~{} -\frac{1}{a^3}\mathcal{D}_t\left[\frac{a^3}{2H}\tP_{,\tX}^2\left(1+\frac{1}{c_s^2}\right)\dot \p_
I \dot \p_J\right]\, ,  
\nonumber 
\label{Interaction matrix}
\end{eqnarray}
where one can substitute the explicit DBI Lagrangian. 

It is worth emphasizing  that, for Lagrangians of the form $P(X,\phi^I)$, the second order action for the perturbations, given in \cite{lr08}, only differs by the coefficient in front of the spatial gradients, which is  ${P}_{,X} G_{IJ}$
(instead of $\tP_{,X} c_s^2 \tG_{IJ}$)  and leads to a different propagation speed along the adiabatic and entropic directions.  By contrast, the DBI Lagrangian (\ref{DD1}-\ref{def_explicit}) gives the same propagation speed for all modes.

We now illustrate the above  formalism in the case where only two scalar fields are present. It is then useful to 
decompose the scalar field perturbations into adiabatic and entropic modes \cite{Gordon:2000hv}, namely
\beq
\label{decomposition}
Q^I=Q_\s e^I+Q_s e^I_s\,,
\eeq
where the entropy vector $e^I_s$ is the  unit vector orthogonal to the adiabatic vector $e^I$, i.e.
\beq
G_{IJ}e_s^I e_s^J=1, \qquad G_{IJ}e_s^I e^J=0.
\eeq
As in standard inflation discussed in the previous section, it is more convenient, after going to conformal time $\tau = \int {dt}/{a(t)}$, to work in terms of the canonically normalized fields
\beq
v_{\s}=\frac{a \sqrt{\tP_X}}{c_s}\, Q_{\s}=\frac{a}{c_s^{3/2}} \, Q_{\s} \,,\qquad \,v_{s}=a\,\sqrt{\tP_X}\, Q_s=\frac{a}{\sqrt{c_s}}\, Q_s\,.
\label{v}
\eeq
 The second order action then becomes
\begin{eqnarray}
\label{S_v}
S_{(2)}&=&\frac{1}{2}\int {\rm d}\tau\,  {\rm d}^3x \Big\{ 
  v_\s^{\prime\, 2}+ v_s^{\prime\, 2} -2\xi v_\s^\prime v_s-c_s^2 \left[(\partial v_\s)^2 + (\partial v_s)^2\right] 
\cr
&& 
\left.
\qquad
+\frac{z''}{z} v_\s^2
+\left(\frac{\alpha''}{\alpha}-a^2 \mu_s^2\right) v_s^2+2\, \frac{z'}{z}\xi v_\s v_s\right\}
\end{eqnarray}
with
\beq
\xi=\frac{a}{\dot \s \tP_{,X} c_s}[(1+c_s^2)\tP_{,s}-c_s^2 \dot\s^2 \tP_{,Xs}]\,, \qquad \dot\s\equiv\sqrt{2X}\, ,
\label{11}
\eeq
and
where we have introduced the two background-dependent  functions 
\beq
z=\frac{a \dot \s }{c_s H}\sqrt{\tP_X}=\frac{a \dot \s }{H c_s^{3/2}}, \qquad \alpha=a\sqrt{\tP_X}=\frac{a}{\sqrt{c_s}}\,.
\eeq
 The effective squared mass $\mu_s^2$ can be computed from the mass matrix (\ref{Interaction matrix}).

The equations of motion derived from the action (\ref{S_v}) can  be written in the compact form
\begin{eqnarray}
v_{\s}''-\xi v_{s}'+\left(k^2c_s^2 -\frac{z''}{z}\right) v_{\s} -\frac{(z \xi)'}{z}v_{s}&=&0\,.
\label{eq_v_sigma}
\\
v_{s}''+\xi  v_{\s}'+\left(k^2 c_s^2- \frac{\alpha''}{\alpha}+a^2\mu_s^2\right) v_{s} - \frac{z'}{z} \xi v_{\s}&=&0\,.
\label{eq_v_s}
\end{eqnarray}
For simplicity, we  assume that  the coupling $\xi$ is very small when the scales of interest cross out the sound horizon, in which case one can quantize the two degrees of freedom independently and solve analytically the system.
The amplification of the vacuum fluctuations at horizon crossing is possible only for very light degrees of freedom. Consequently, if $\mu_s^2$ is larger than $H^2$, this amplification is suppressed and there is no production of entropy modes. From now on,  we assume that $|\mu_s^2|/H^2\ll 1$.

Following the standard procedure outlined in the previous section, 
  one selects the  positive frequency solutions of Eqs.~(\ref{eq_v_sigma}) and (\ref{eq_v_s}), which  correspond to the usual vacuum on very small scales:
  \beq
v_{\s\, k} \simeq v_{s\, k} \simeq  \frac{1}{\sqrt{2k c_s}}e^{-ik c_s \tau }\left(1-\frac{i}{k c_s\tau}\right)\, .
\eeq
As a consequence, the power spectra for $v_\s$ and $v_s$ after sound horizon crossing  have the same amplitude\beq
{\cal P}_{v_\s}={\cal P}_{v_s}=\frac{k^3}{2\pi^2}|v_{\s\, k}|^2\simeq\frac{H^2 a^2}{4\pi^2 c_s^3}.
\eeq
 However, in terms of the initial fields $Q_\s$ and $Q_s$, one finds, using (\ref{v}), 
\beq
\label{power_sigma}
{\cal P}_{Q_\s*}\simeq\frac{H^2}{4\pi^2 }, \quad {\cal P}_{Q_s*}\simeq\frac{H^2}{4\pi^2 c_s^2},
\eeq
(the subscript $*$ indicates that the corresponding quantity is evaluated at sound horizon crossing $k c_s=aH$)
which shows that, for small $c_s$, the entropic modes are {\it amplified} with respect to the adiabatic modes:
\beq
Q_{s*}\simeq \frac{Q_{\sigma*}}{c_s}.
\eeq

In order to confront the predictions of inflationary models to cosmological observations, it is useful to rewrite the scalar field perturbations in terms of geometrical quantities.
The comoving curvature perturbation is related to the adiabatic perturbation by the expression 
 \beq
{\cal R}=\left( \frac{H}{2 P_{<IJ>}X^{IJ}} \right) P_{<KL>} \dot{\phi}^K Q^L=\frac{H}{\dot \s}Q_{\s}\,.
\label{R}
\eeq
One thus recovers the usual {\it single-field} result \cite{Garriga:1999vw} that the power spectrum for $\R$ at sound horizon crossing is given by
\beq
{\cal P}_{\cal R_*}=\frac{k^3}{2\pi^2}\frac{|v_{\s\, k}|^2}{z^2}\simeq\frac{H^4}{4\pi^2  \dot\s^2 }=\frac{H^2}{8\pi^2 \epsilon c_s }\,,
\label{power-spectrum-R}
\eeq
where $\epsilon=-\dot H / H^2\,$.  It is then convenient to define an entropy perturbation, which we denote ${\cal S}$, such that its power spectrum
at sound horizon crossing is the same as that of the curvature perturbation,
\beq
{\cal S}=c_s\frac{H}{\dot \s}Q_{s}\, ,
\label{S}
\eeq
so that 
\beq
{\cal P}_{\cal S_*}={\cal P}_{\cal R_*}\equiv {\cal P}_{*}.
\eeq

The power spectrum for the tensor modes is still governed by the transition at {\it Hubble radius} and its amplitude, given by 
\beq
{\cal P}_{\cal T}=\left(\frac{2H^2}{\pi^2}\right)_{k=aH}\,,
\label{power-spectrum-T}
\eeq
is much smaller than the curvature amplitude in the small $c_s$ limit.

Leaving aside the possibility that the entropy modes during inflation lead  to primordial entropy fluctuations {\it after} inflation that could be directly detectable in the CMB fluctuations (potentially correlated with adiabatic modes as first discussed 
in \cite{Langlois:1999dw}), we consider here only the influence of the entropy modes on the final curvature perturbation. 
In contrast with the single-field case, the curvature 
perturbation generally evolves on large scales in a multi-field scenario \cite{sy} (see also \cite{Lalak:2007vi} for a recent analysis with non-standard kinetic terms), because of the entropy modes. This transfer from the entropic to the adiabatic modes depends on the details of the scenario and of the background trajectory in field space but it can be parametrized by a transfer coefficient  
which appears in the formal solution $\R=\R_*+T_{ {\cal R}  {\cal S} } \cal S_*$ of the evolution equations, which are of the form
\beq
\dot {\cal R}\approx\alpha H {\cal S}  \qquad \dot {\cal S}\approx\beta H {\cal S}
\label{R-S-evolution}\,.
\eeq
This implies in particular that the final curvature power-spectrum can be formally expressed as 
\beq
{\cal P}_{\cal R}=(1+T_{{\cal R} {\cal S}}^2) {\cal P}_{\cal R_{*}}
\label{observed-spectrum}
\eeq
It is sometimes useful to define the ``transfer angle'' $\Theta$ ($\Theta=0$ if there is no transfer and $|\Theta|=\pi/2$ if the final curvature perturbation is mostly of entropic origin) by
\beq
{\sin} \Theta =\frac{T_{ {\cal R}  {\cal S} }}{\sqrt{1+T^2_{ {\cal R}  {\cal S} }}}\,.
\label{correlation-result}
\eeq
The relation between the curvature power-spectrum at sound-horizon crossing and  its observed value is thus
\beq
{\cal P}_{\cal R_{*}}={\cal P}_{\cal R} {\rm cos^2} \Theta
\label{PR-initial}
\eeq
Using the tensor amplitude Eq.~(\ref{power-spectrum-T}), one finds that  the tensor to scalar ratio is given by
\beq
r \equiv \frac{{\cal P}_{\cal T}}{{\cal P}_{\cal R}}=16\, \epsilon\, c_s {\cos^2} \Theta.
\label{r}
\eeq
Interestingly this expression combines the result of $k$-inflation, where the ratio is suppressed by the sound speed $c_s$, and that of standard multi-field inflation.

It is also possible to compute the non-Gaussianities generated in these models \cite{lrst08a, lrst08b} (see also 
\cite{Gao:2008dt} and \cite{Arroja:2008yy}). For multi-field 
DBI inflation described by the Lagrangian (\ref{DD1}), we find that the shape of  non-Gaussianities is the same as in single-field DBI but their amplitude is affected by the transfer between the entropic and adiavatic modes.  The contribution from the scalar field three-point functions  to the coefficient $f_{NL}$ is given by
\beq
f_{NL}^{(3)}=-\frac{35}{108}\frac{1}{c_s^2}\frac{1}{1+T^2_{{\cal R} {\cal S}} }=-\frac{35}{108}\frac{1}{c_s^2} {\cos^2} \Theta \,.
\label{f_NL3}
\eeq 
This result is the consequence  of the amplication of both the power spectrum and
the three-point correlation function of ${\cal R}$
by a factor $(1+T^2_{{\cal R} {\cal S}})$. Since $f_{NL}$ is roughly the ratio of the three-point function with respect to the {\it square} of the power spectrum, this implies that $f_{NL}$ is  {\it reduced} by the factor 
$(1+T^2_{{\cal R} {\cal S}})$. The effect of entropy modes is therefore potentially important in the perspective 
of confronting DBI models to  future CMB observations.

\section{Conclusion}
\label{sec:conclusion}
In this contribution, we have presented  a general analysis of cosmological perturbations in multi-field inflationary models, allowing for non standard kinetic terms. This approach generalizes much more restrictive situations  considered previously in the literature and is motivated by recent constructions of inflationary models in the context of string theory, where multiple fields and non standard kinetic terms are very common, a typical example being DBI inflation. 

As we have tried to show, multi-field inflation is a very rich playground, where entropy modes can play a significant role. The most important consequence of entropy modes is the possibility to  modify the curvature perturbation, on large scales, in contrast with single field inflation. This means that the adiabatic fluctuations, which we observe today in the CMB, could come originally from entropy perturbations produced during multi-field inflation. 

\vspace{0.3cm}
{\bf Acknowledgment:} I would like to thank the organizers for inviting me to a very friendly and stimulating conference.

\end{document}